\documentstyle[epsfig,epsf,twoside]{article}

\catcode`\@=11
\long\def\@makefntext#1{
\protect\noindent \hbox to 3.2pt {\hskip-.9pt  
$^{{\eightrm\@thefnmark}}$\hfil}#1\hfill}		%CAN BE USED 

\def\thefootnote{\fnsymbol{footnote}}
\def\@makefnmark{\hbox to 0pt{$^{\@thefnmark}$\hss}}	%ORIGINAL 
	
\def\ps@myheadings{\let\@mkboth\@gobbletwo
\def\@oddhead{\hbox{}
\rightmark\hfil\eightrm\thepage}   
\def\@oddfoot{}\def\@evenhead{\eightrm\thepage\hfil
\leftmark\hbox{}}\def\@evenfoot{}
\def\sectionmark##1{}\def\subsectionmark##1{}}

\oddsidemargin=\evensidemargin
\renewcommand{\thefootnote}{\fnsymbol{footnote}}

\newcounter{sectionc}\newcounter{subsectionc}\newcounter{subsubsectionc}
\renewcommand{\section}[1] {\vspace{12pt}\addtocounter{sectionc}{1} 
\setcounter{subsectionc}{0}\setcounter{subsubsectionc}{0}\noindent 
	{\tenbf\thesectionc. #1}\par\vspace{5pt}}
\renewcommand{\subsection}[1] {\vspace{12pt}\addtocounter{subsectionc}{1} 
	\setcounter{subsubsectionc}{0}\noindent 
	{\bf\thesectionc.\thesubsectionc. {\kern1pt \bfit #1}}\par\vspace{5pt}}
\renewcommand{\subsubsection}[1] {\vspace{12pt}\addtocounter{subsubsectionc}{1}
	\noindent{\tenrm\thesectionc.\thesubsectionc.\thesubsubsectionc.
	{\kern1pt \tenit #1}}\par\vspace{5pt}}
\newcommand{\nonumsection}[1] {\vspace{12pt}\noindent{\tenbf #1}
	\par\vspace{5pt}}

\newcounter{appendixc}
\newcounter{subappendixc}[appendixc]
\newcounter{subsubappendixc}[subappendixc]
\renewcommand{\thesubappendixc}{\Alph{appendixc}.\arabic{subappendixc}}
\renewcommand{\thesubsubappendixc}
	{\Alph{appendixc}.\arabic{subappendixc}.\arabic{subsubappendixc}}

\renewcommand{\appendix}[1] {\vspace{12pt}
        \refstepcounter{appendixc}
        \setcounter{figure}{0}
        \setcounter{table}{0}
        \setcounter{lemma}{0}
        \setcounter{theorem}{0}
        \setcounter{corollary}{0}
        \setcounter{definition}{0}
        \setcounter{equation}{0}
        \renewcommand{\thefigure}{\Alph{appendixc}.\arabic{figure}}
        \renewcommand{\thetable}{\Alph{appendixc}.\arabic{table}}
        \renewcommand{\theappendixc}{\Alph{appendixc}}
        \renewcommand{\thelemma}{\Alph{appendixc}.\arabic{lemma}}
        \renewcommand{\thetheorem}{\Alph{appendixc}.\arabic{theorem}}
        \renewcommand{\thedefinition}{\Alph{appendixc}.\arabic{definition}}
        \renewcommand{\thecorollary}{\Alph{appendixc}.\arabic{corollary}}
        \renewcommand{\theequation}{\Alph{appendixc}.\arabic{equation}}
        \noindent{\tenbf Appendix \theappendixc #1}\par\vspace{5pt}}
\newcommand{\subappendix}[1] {\vspace{12pt}
        \refstepcounter{subappendixc}
        \noindent{\bf Appendix \thesubappendixc. {\kern1pt \bfit #1}}
	\par\vspace{5pt}}
\newcommand{\subsubappendix}[1] {\vspace{12pt}
        \refstepcounter{subsubappendixc}
        \noindent{\rm Appendix \thesubsubappendixc. {\kern1pt \tenit #1}}
	\par\vspace{5pt}}

\newcommand{\textlineskip}{\baselineskip=13pt}
\newcommand{\smalllineskip}{\baselineskip=10pt}

\def\eightcirc{
\begin{picture}(0,0)
\put(4.4,1.8){\circle{6.5}}
\end{picture}}
\def\eightcopyright{\eightcirc\kern2.7pt\hbox{\eightrm c}} 

\newcommand{\copyrightheading}[1]
	{\vspace*{-2.5cm}\smalllineskip{\flushleft
	{\footnotesize International Journal of Modern Physics A, #1}\\
	{\footnotesize $\eightcopyright$\, World Scientific Publishing
	 Company}\\
	 }}

\newcommand{\publisher}[2]{{\begin{center}\footnotesize\smalllineskip 
	Received #1\\
	Revised #2
	\end{center}
	}}

\def\abstracts#1#2#3{{
	\centering{\begin{minipage}{4.5in}\baselineskip=10pt\footnotesize
	\parindent=0pt #1\par 
	\parindent=15pt #2\par
	\parindent=15pt #3
	\end{minipage}}\par}} 

\renewenvironment{thebibliography}[1]
	{\frenchspacing
	 \ninerm\baselineskip=11pt
	 \begin{list}{\arabic{enumi}.}
	{\usecounter{enumi}\setlength{\parsep}{0pt}
	 \setlength{\leftmargin 12.7pt}{\rightmargin 0pt} %FOR 1--9 ITEMS
	 \setlength{\itemsep}{0pt} \settowidth
	{\labelwidth}{#1.}\sloppy}}{\end{list}}

\newcounter{itemlistc}
\newcounter{romanlistc}
\newcounter{alphlistc}
\newcounter{arabiclistc}

\newcommand{\fcaption}[1]{
        \refstepcounter{figure}
        \setbox\@tempboxa = \hbox{\footnotesize Fig.~\thefigure. #1}
        \ifdim \wd\@tempboxa > 5in
           {\begin{center}
        \parbox{5in}{\footnotesize\smalllineskip Fig.~\thefigure. #1}
            \end{center}}
        \else
             {\begin{center}
             {\footnotesize Fig.~\thefigure. #1}
              \end{center}}
        \fi}

\newcommand{\tcaption}[1]{
        \refstepcounter{table}
        \setbox\@tempboxa = \hbox{\footnotesize Table~\thetable. #1}
        \ifdim \wd\@tempboxa > 5in
           {\begin{center}
        \parbox{5in}{\footnotesize\smalllineskip Table~\thetable. #1}
            \end{center}}
        \else
             {\begin{center}
             {\footnotesize Table~\thetable. #1}
              \end{center}}
        \fi}

\def\@citex[#1]#2{\if@filesw\immediate\write\@auxout
	{\string\citation{#2}}\fi
\def\@citea{}\@cite{\@for\@citeb:=#2\do
	{\@citea\def\@citea{,}\@ifundefined
	{b@\@citeb}{{\bf ?}\@warning
	{Citation `\@citeb' on page \thepage \space undefined}}
	{\csname b@\@citeb\endcsname}}}{#1}}

\newif\if@cghi
\def\cite{\@cghitrue\@ifnextchar [{\@tempswatrue
	\@citex}{\@tempswafalse\@citex[]}}
\def\citelow{\@cghifalse\@ifnextchar [{\@tempswatrue
	\@citex}{\@tempswafalse\@citex[]}}
\def\@cite#1#2{{$\null^{#1}$\if@tempswa\typeout
	{IJCGA warning: optional citation argument 
	ignored: `#2'} \fi}}

\def\pmb#1{\setbox0=\hbox{#1}
	\kern-.025em\copy0\kern-\wd0
	\kern.05em\copy0\kern-\wd0
	\kern-.025em\raise.0433em\box0}

\def\fnt#1#2{\footnotetext{\kern-.3em
	{$^{\mbox{\scriptsize #1}}$}{#2}}}

\def\fpage#1{\begingroup
\voffset=.3in
\thispagestyle{empty}\begin{table}[b]\centerline{\footnotesize #1}
	\end{table}\endgroup}

\def\runninghead#1#2{\pagestyle{myheadings}
\markboth{{\protect\footnotesize\it{\quad #1}}\hfill}
{\hfill{\protect\footnotesize\it{#2\quad}}}}
\font\tenrm=cmr10
\font\tenit=cmti10 
\font\tenbf=cmbx10
\font\bfit=cmbxti10 at 10pt
\font\ninerm=cmr9

\font\eightrm=cmr8

\def\qed{\hbox{${\vcenter{\vbox{			%HOLLOW SQUARE
   \hrule height 0.4pt\hbox{\vrule width 0.4pt height 6pt
   \kern5pt\vrule width 0.4pt}\hrule height 0.4pt}}}$}}

\renewcommand{\thefootnote}{\fnsymbol{footnote}}	%USE SYMBOLIC FOOTNOTE

\def\lsim{\mathrel{\lower4pt\hbox{$\sim$}}\hskip-12.5pt\raise1.6pt\hbox{$<$}\;}
\def\gsim{\mathrel{\lower4pt\hbox{$\sim$}}\hskip-12.5pt\raise1.6pt\hbox{$>$}\;}

\begin{document}

\runninghead{Aspects of parity, CP, and time reversal violation in hot QCD}
 {D.E. Kharzeev, R.D. Pisarski and M.H.G. Tytgat}

\normalsize\textlineskip
\thispagestyle{empty}
\setcounter{page}{1}

\copyrightheading{}			%{Vol. 0, No. 0 (1993) 000--000}

\vspace*{0.88truein}

\fpage{1}
\centerline{\bf ASPECTS OF PARITY, CP, AND TIME REVERSAL}
\vspace*{0.035truein}
\centerline{\bf VIOLATION IN HOT QCD\footnote{Talk given at SEWM2000, Marseille, June 14-17 2000 and at ISMD2000, Tihany, October 9-15 2000.}
%MANUSCRIPTS USING COMPUTER SOFTWARE\footnote{For
%the title, try not to use more than 3 lines. Typeset the title
%in 10 pt Times Roman, uppercase and boldface.}
}
\vspace*{0.37truein}
\centerline{\footnotesize DMITRI E. KHARZEEV, ROBERT D. PISARSKI}
%\footnote{Typeset names in
%10 pt Times Roman, uppercase. Use the footnote to indicate the
%present or permanent address of the author.}

\vspace*{0.015truein}
\centerline{\footnotesize\it Physics Department}
\baselineskip=10pt
\centerline{\footnotesize\it Brookhaven National Laboratory
}
\centerline{\footnotesize\it Upton, New York 11973-5000, USA
}
\vspace*{10pt}
\centerline{\footnotesize MICHEL H.G. TYTGAT}
\vspace*{0.015truein}
\centerline{\footnotesize\it Physique Th\'eorique, CP225, Universit\'e Libre de Bruxelles}
\baselineskip=10pt
\centerline{\footnotesize\it Bld du Triomphe, 1050 Bruxelles, Belgium}

\vspace*{0.225truein}
\publisher{(received date)}{(revised date)}

\vspace*{0.21truein}
\abstracts{
We discuss various aspects of parity, CP, and time reversal 
invariances in QCD. 
In particular, we focus attention on the previously proposed 
possibility that 
these experimentally established symmetries of strong interactions 
may be broken 
at finite temperature and/or density. This   
would have dramatic signatures in relativistic heavy ion collisions; 
we describe 
some of the most promising signals. 
}{}{}

%\textlineskip			%) USE THIS MEASUREMENT WHEN THERE IS
%\vspace*{12pt}			%) NO SECTION HEADING

\vspace*{1pt}\textlineskip	%) USE THIS MEASUREMENT WHEN THERE IS
\section{Introduction}	%) A SECTION HEADING
\vspace*{-0.5pt}
\noindent
It is commonly accepted that strong interactions do not break parity or CP.
This is because in QCD, CP violation can arise only through a 
non-zero value of 
the $\theta$ angle, and the experimental limit \cite{edm} 
on the  neutron dipole 
electric moment imposes a stringent limit $\theta \lsim 3\ 10^{-10}$. 
Why $\theta$ is so close to zero in Nature is a fundamental 
question, known as 
the ``Strong CP Problem''. There are various proposed 
solutions to this problem, the most popular being the axion 
hypothesis (see \cite{axion}), 
but the discussion of these solutions is not the 
topic of this talk. Instead we shall assume that, for whatever 
reason, $\theta=0$ precisely, 
and discuss the behavior of the theory at finite temperature and/or density 
\cite{Kharzeev:1998kz}.   

If $\theta=0$, a theorem by Vafa and Witten states that the 
vacuum is unique and 
conserves CP~\cite{Vafa:1984xg}.
There are circumstances under which the Vafa-Witten theorem does not apply, 
however. 
For instance, no conclusion can be drawn in the presence of a finite net  
density of quarks. 
Indeed, the existence of degenerate vacuum states with opposite parity in the 
superconducting phase of QCD has been discussed by Pisarski 
and Rischke~\cite{Pisarski:1999dk}, 
and the (pseudoscalar) pion condensate at finite 
isospin density has been found by Son and Stephanov~\cite{SonSt}.  
If $\theta=\pi$, the theorem also does not apply. 
{\em A priori}, this corresponds to a CP conserving case because 
 $\theta \rightarrow - \theta$, $\pi$ and $-\pi$ are 
equivalent modulo $2 \pi$. However, 
CP can still be spontaneously
broken, a phenomenon discovered by Dashen before the 
advent of QCD~\cite{Dashen:1971et}.
Finally, the theorem is only a statement about the ground state of the theory. 
In particular, it does 
not preclude the existence of metastable 
vacua within which CP is broken. In this talk we gather some of the evidence
for the existence of extra vacua in hot QCD, and discuss its
possible relevance to the ongoing relativistic heavy ion program. 

%\pagebreak

\textheight=7.8truein
\setcounter{footnote}{0}
\renewcommand{\thefootnote}{\alph{footnote}}

\section{$\Theta$ vacuum in gluodynamics}
\noindent
Consider the Euclidean Yang-Mills action with  gauge group $SU(N_c)$,
\begin{equation}
S = {1\over 4 g^2} \int\mbox{\rm Tr} F^2 + 
{\theta \over 32 \pi^2} \int \!d^4\!x \ \mbox{\rm Tr} F\tilde F 
\end{equation}
where $\tilde F_{\mu\nu} = {1\over 2} 
\epsilon_{\mu\nu\alpha\beta} F_{\alpha \beta}$. Because 
the integrand in the second term is 
a total derivative $q(x) = 1/32 \pi^2 \ 
\mbox{\rm Tr} F \tilde F = \partial_\mu K_\mu$, it 
cannot influence equations of motion on the classical level, and all $\theta$ 
dependence is through the quantum anomaly. 
In the dilute instanton gas approximation, the ground
 state energy is 
\begin{equation}
E(\theta) \propto e^{- 8\pi^2/ g^2} \cos\theta,
\end{equation}
which illustrates that $\theta$ is an angle.

While the dilute instanton gas should be an adequate 
concept at high temperature, 
when the instantons are suppressed by screening,  
we would like to rely on another approximation 
in the confined phase. 
Unfortunately, lattice simulations are 
inefficient for $\theta \neq 0$, and the only 
reliable analytical limit of the theory 
we know of is the case of large $N_c$. 
The large $N_c$ limit \cite{largeN} is reached by rescaling the gauge coupling 
$g^2 \rightarrow \lambda = g^2 N_c$, with $\lambda$ 
kept fixed as $N_c \rightarrow \infty$. 
Similarly, one must rescale $\theta$ to $\theta/N_c$. The latter implies that 
the low energy effective action for YM theory in the 
confined phase takes the form
\begin{equation}
{\cal L}_{\rm eff} = N_c^2\ {\cal L}[\Phi,\partial \Phi, \ldots, \theta/N_c]
\end{equation} 
where $\Phi$ denotes some generic colorless (glueball) state, with mass 
$M_\Phi = {\cal O}(N_c^0)$. The $N_c^2$
 factor comes from counting the number of degrees of 
freedom.  In particular the vacuum energy as function of $\theta$ is
\begin{equation}
E[\theta] = N_c^2\ F[\theta/N_c]
\end{equation}
To leading order at large $N_c$, 
\begin{equation}
E[\theta] \approx N_c^2 F[0] + {1\over 2} \chi\ \theta^2
\end{equation}
where $\chi= \int d^4\!x \ \langle q(x) q(0) \rangle\sim {\cal O}(N_c^0)$ 
is the topological susceptibility. There is a little problem however:
this expression is manifestly incompatible with the 
requirement of $E[\theta] \equiv E[\theta + 2 \pi]$. 
The remedy has been proposed by Witten~\cite{Witten:1998uk}. If one
takes
\begin{equation}
\label{vacen}
E[\theta] = N_c^2 \min_k F\left ({\theta + 2 k \pi\over N_c}\right)
\end{equation}
the vacuum energy becomes a multivalued function, 
with $k$ labelling the different branches.
A similar construction is known to 
arise, for instance, in the Schwinger model~\cite{Coleman:1976uz}. 
\begin{figure}[htp]
\vspace*{10pt}
\centerline{
\epsfig{file=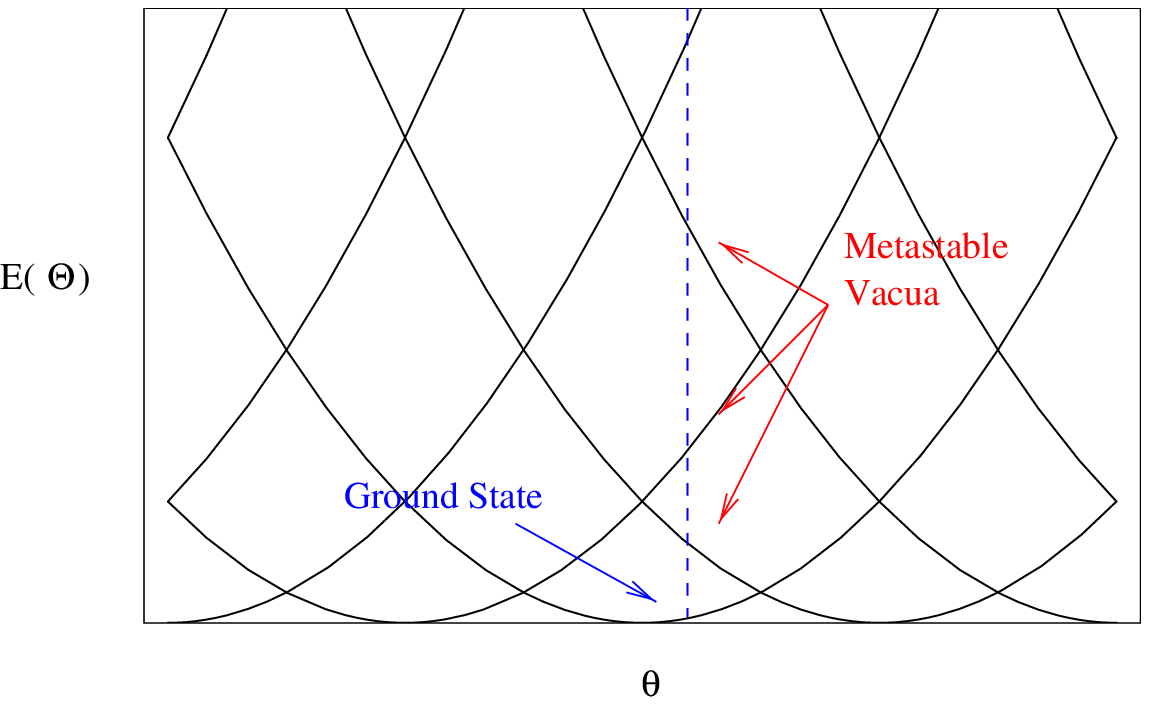,height=4cm}		
%ORIGINAL SIZE=1.6TRUEIN x 100% - 0.2TRUEIN
}
\vspace*{5pt}
\fcaption{Vacuum energy in gluodynamics as a function of 
$\theta$. At fixed $\theta$, there is a ground state and a tower of extra vacua.}
\end{figure}
The expression ~(\ref{vacen}) is quite remarkable: 
it predicts that for each fixed $\theta$, there is a tower of extra
 vacua (see Fig.1). 

A corollary is that at $\theta=\pi \bmod 2 \pi$
 there are two degenerate vacua and CP is spontaneously broken. 
Also, there is a stable domain wall in the spectrum, very 
much like in the Schwinger model.  If $\theta$ departs from $\pi$, one of the vacua becomes unstable. The decay rate of this 
false vacuum has been estimated to scale with the number of colors as 
 $\Gamma \sim \exp( - N_c^4)$
~\cite{Witten:1998uk,Shifman:1999if,Gabadadze:1999na,Tytgat:1999yx} 
-- as $N_c\rightarrow \infty$, the 
metastable (lowest lying) vacua become stable.

\section{Vacuum in QCD with $N_f$ light quark flavors}
\noindent
The low energy phenomenological Lagrangian that 
incorporates both leading order quark
mass and large $N_c$ effects 
is~\cite{DiVecchia:1980ve,Witten:1980sp,Rosenzweig:1980ay,Nath:1981ik}
\begin{equation}
\label{dvw}
{\cal L} = {f_\pi^2\over  4} \, \mbox{\rm tr}( \partial_\mu U^\dagger 
\partial^\mu U ) + \Sigma \, \mbox{\rm Re}\left[
\mbox{\rm tr} ({\cal M}\,U^\dagger)\right] 
-{\chi \over 2} (\theta + i\, \log \det U)^2
\end{equation}
where ${\cal M} = \mbox{\rm diag}(m_u,m_d,m_s)$ is the quark  mass
matrix, $\Sigma = \vert \langle  \bar q q\rangle\vert$, 
 $U = \exp i \phi^a \lambda^a/f_\pi$ is a $N_f\times N_f$ unitary 
matrix and the $\phi^a$ are the $a=1,\ldots,N_c^2$ would be Goldstone modes. 
Specialising to the neutral modes, the potential term 
in~(\ref{dvw}) gives
\begin{equation}
\label{pot}
{\cal V} = - \sum_{i=u,d,s} m_i \Sigma \cos{\phi_i\over f_\pi} + 
{\chi\over 2} (\theta - \sum_{i=u,d,s} \phi_i/f_\pi)^2
\end{equation} 
A non-vanishing topolological susceptibility 
explicitly breaks the $U(1)_A$ symmetry and 
gives a mass to the $\eta^\prime$ meson, 
$m^2_{\eta^\prime} \approx \chi/f_\pi^2$   
\cite{Veneziano:1979ec,Witten:1979vv}. 
In Nature, the value of the susceptibility in the vacuum is  
$\chi \approx (180\  \mbox{\rm MeV})^4$. In one plots
 the potential for realistic values of the quark mass, 
the result is a set of almost perfect parabolic curves : this is because the 
$\theta$ term dominates over the terms containing the quark masses.
\footnote{Some interesting features arise at $\theta \sim \pi$; these include 
Dashen's phenomenon and the appearance of stable domain walls, but 
unfortunately this is beyond the scope of this 
talk~\cite{Creutz:1995wf,Smilga:1999dh}.}
 In the large $N_c$ limit, because 
$\Sigma \propto N_c$, things become more interesting. 
To simplify the discussion, let us 
concentrate on the case of one light quark flavor, with pseudo-Goldstone field 
$\eta$. For $\theta = 0$, the potential 
energy at large $N_c$ is as in Fig.2. 
There is an absolute minimum at $\eta= 0$ 
and extra local minima at $\eta \approx 2 \pi \ k$. 
These extra minima violate $CP$ because
\begin{equation}
\langle q \rangle = 
{\partial {\cal V}\over\partial \theta} \approx \chi\ \langle 
\eta\rangle \stackrel{CP}{\longrightarrow} - \chi\ \langle \eta \rangle  
\end{equation}
\begin{figure}[htp]
\vspace*{10pt}
\centerline{
\epsfig{file=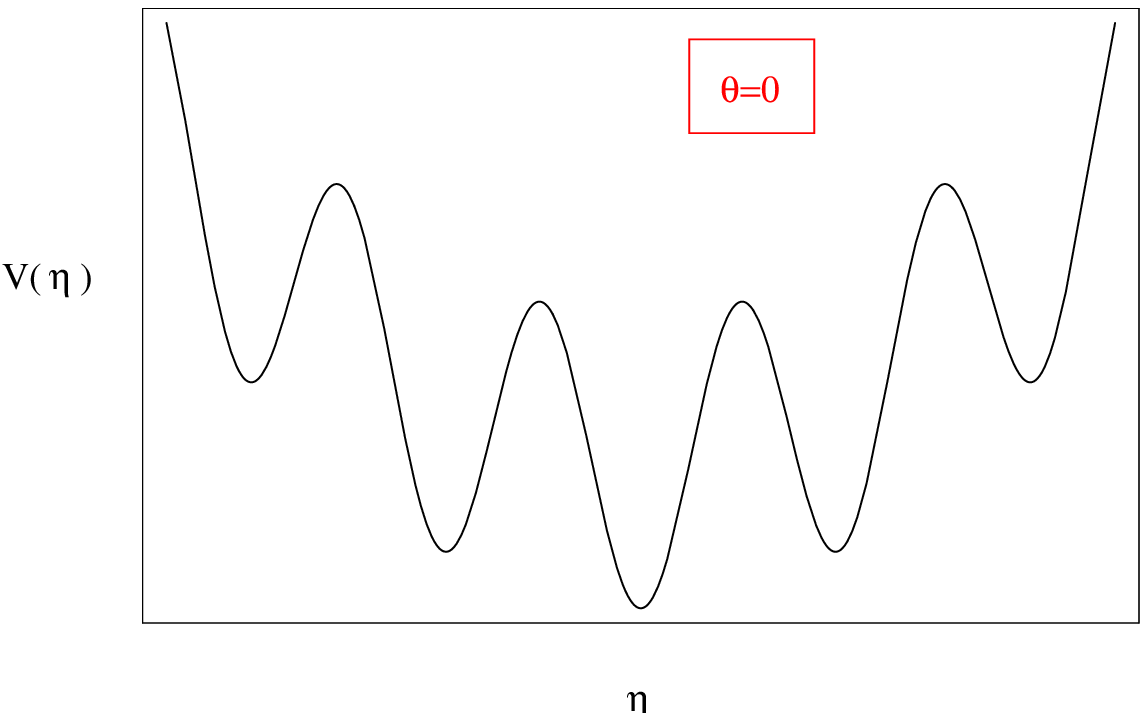,height=4cm}		
%ORIGINAL SIZE=1.6TRUEIN x 100% - 0.2TRUEIN
}
\vspace*{5pt}
\fcaption{Potential for $\eta$ at large $N_c$ and $\theta=0$.}
\end{figure}
A technical remark is in order here. The potential~(\ref{pot}) is not periodic 
in the flavor singlet phase. On the other hand it is 
periodic in $\theta$. Specialising
again to one light quark flavor, we see that
a shift $\theta \rightarrow \theta + 2 \pi$ is 
compensated by $\eta \rightarrow \eta - 2 \pi$. 
Incidentally, this remark provides us with a way 
to derive the pure Yang-Mills potential introduced in
the previous section. In the large $N_c$ limit, one can 
integrate out the $\eta$ field to the 
\begin{equation}
E[\theta] \approx {1\over 2} \chi\ \min_k (\theta + k\ \langle \eta \rangle)^2 
\approx {1\over 2} \chi\ \min_k (\theta + k\ 2 \pi)^2
\end{equation}
This construction is {\em ad hoc} as the $\eta$ field is essentially made of 
 quark-antiquark pairs while the pure YM potential should be made of gluons. 
This is remedied if we postulate the existence of a 
heavy pseudoscalar ``glueball''   
field $\Omega$~\cite{Me} and consider instead the following 
modified $N_f=1$ potential   
\begin{equation}
\label{ampot}
{\cal V}^\prime = - \Lambda^4 \cos \Omega - m \Sigma \cos \eta + {1\over 2} 
\chi (\theta + \Omega + \eta)^2
\end{equation}
This potential is now manifestly  $2\pi$-periodic  
in $\theta$ {\em and}  
$\eta$ and $\Omega$. The fact that understanding the $\theta$ 
dependence requires 
the introduction of a gluonic field is perhaps not surprising -- after all, 
the physical $\eta'$ field must contain a  significant gluonic component.  
Similar potentials, with extra heavy degrees of freedom, have been 
suggested to arise in some supersymmetric YM 
theories, like ${\cal N}=1$ 
SYM~\cite{Kogan:1998dt,Gabadadze:1999bi,Gabadadze:2000pp}. Various arguments
indicate that these degree of freedom
 should become very heavy in the large $N_c$ limit, $M_\Omega \propto N_c$.\cite{Gabadadze:2000pp}. This is unlike the usual glueball states, whose mass is ${\cal O}(N_c)$.
If we integrate out the heavy $\Omega$ field, the effective potential becomes
multivalued, pretty much like the pure Yang-Mills potential of 
Eq.~(\ref{vacen}). Such a potential has 
been constructed in~\cite{Halperin:1998rc}, 
starting from the QCD partition function.

\begin{figure}[htp]
\vspace*{10pt}
\centerline{
\epsfig{file=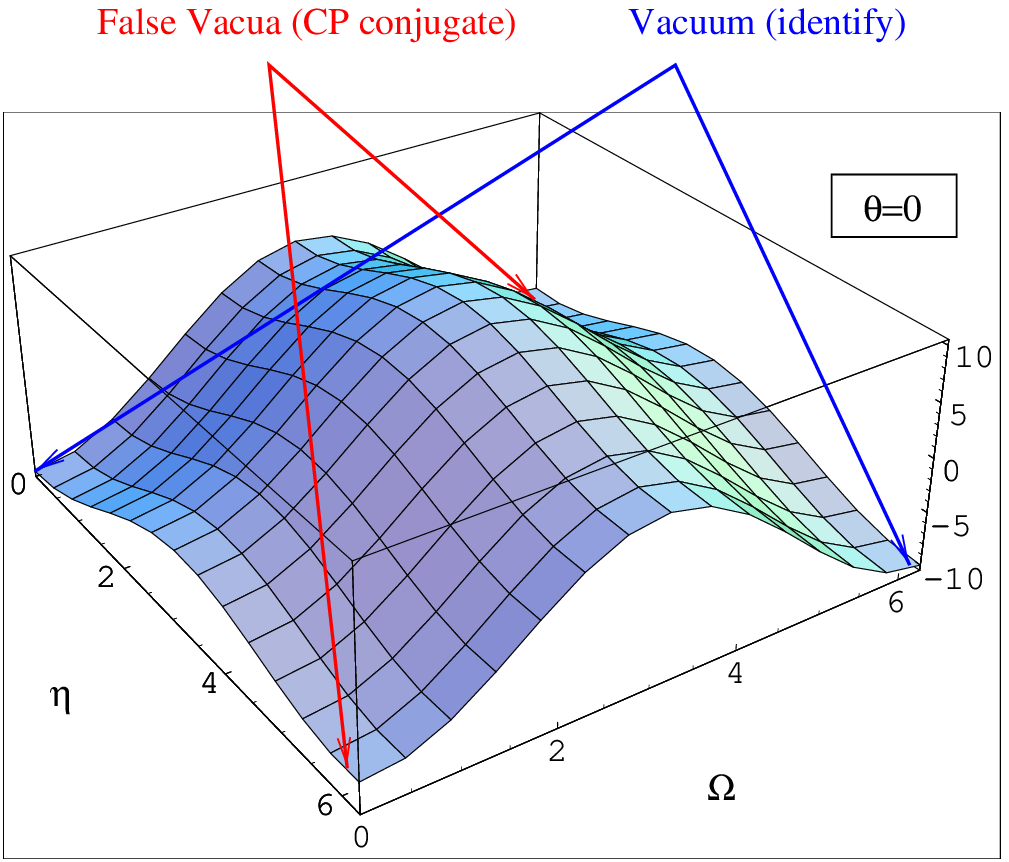,height=6cm}		
%ORIGINAL SIZE=1.6TRUEIN x 100% - 0.2TRUEIN
}
\vspace*{5pt}
\fcaption{Amended potential for $\eta$ at $\theta=0$. The relative scales have been chosen so as to give a pleasant picture.}
\end{figure}

\section{$U(1)_A$ symmetry at finite temperature and density}
\noindent

The considerations of the last section should 
apply in the limit of very large number of 
colors, $N_c \rightarrow \infty$. Inversely, for fixed $N_c$, 
the same effects would arise if the topological susceptibility
 $\chi$ drops with respect  to its vacuum expectation value,
 $\chi \sim (180 \ \mbox{\rm MeV})^2$~\cite{Kharzeev:1998kz} 
or, in other words, if there is 
an approximate restoration of $U(1)_A$ symmetry.
This could arise at finite temperature and/or density. 
Let us consider again  the limit of
large number of colors. At very high temperature, because of
Debye screening 
in the deconfined phase of QCD, the
instanton calculus is reliable.  Contributions to the topological 
susceptibility are then exponentially small 
$\chi\sim \exp(- N_c/\lambda) \sim 0$. On the other hand, in
the confined phase, $\chi \sim {\cal O}(1)$. The transition 
between these two behaviors is not well understood. 
Toy model calculations in $D=1+1$ 
show that
the instanton picture can be extended all the way down to the critical 
temperature, but not beyond~\cite{D'Adda:1978uc,Affleck:1980gy,Davis:1986qw}.
\footnote{In $D=3 + 1$, the AdS/CFT has shed some light on this
issue. Instanton calculus is applicable in theories 
with conformal invariance, as  
${\cal N}=4$ SYM~\cite{Banks:1998nr}, or when supersymmetry allows to 
holomorphically interpolate 
between weak and strong couplings. In non-supersymmetric theories, however, 
there seems to be a drastic transition between the instanton and the 
large $N_c$ description~\cite{Barbon:1999zp}.} In $D=3+1$, lattice 
simulations of the topological susceptibility show a similar behavior: 
the topological susceptibility is 
essentially constant at low temperature, and 
then drops sharply near $T_c$.~\cite{Alles:1999mt,deForcrand:1997ut}. 
Whether this effect permits the appearance of metastable vacua is
hard to answer. 
If we nevertheless naively extrapolate the range of validity 
of the effective potential~(\ref{dvw})
to higher temperature, we see that $\chi$ must drop dramatically
because the existence of the extra vacua is controlled by the
mass of the lightest quarks, 
$\chi(T_c)/\chi(0) \sim 5\%$~\cite{Kharzeev:1998kz} 
\footnote{This is one of the manifestations of the
fact there can be no CP
violation if any of the quark is massless.}.
While the current lattice simulations
do not seem to see such a large suppression, 
they are not far from it. Also, things could 
improve at finite $T$ and $\mu$, where $\mu$ is the quark chemical potential.
The argument is as follows.

If the $U(1)_A$ symmetry is effectively restored 
at $T_c$, then universality arguments 
suggest that the chiral phase transition should be  first order, driven by 
fluctuations~\cite{Pisarski:1984ms}. Consider now the 
phase diagram for QCD with two light quark flavors 
in the plane of temperature and chemical potential, 
$T-\mu$. Even though there is no
phase transition along the $\mu=0$ axis, various arguments suggests
that there is  a line of first order phase transitions, terminated by a
critical point $(T_\ast,\mu_\ast)$ \cite{Hsu:1998eu}. 
The position of the point is  uncertain, since it is 
influenced by various effects. 
For instance if the mass of the strange quark 
were below some critical value $m_s^\ast$, the critical 
point would move toward the temperature
axis and the chiral symmetry breaking phase transition would be
first order everywhere.
A similar conclusion would be reached if the $U(1)_A$ symmetry was restored. 
This could 
be controlled by changing the number of colors, for instance \footnote{
Ideally one would prefer a dynamical explanation, such as the one 
provided by the 
instanton liquid picture.\cite{Schafer:1998wv} }.
The fact that $N_c$ is a discrete  
variable is not really an issue here. What we want to
emphasize is that, perhaps, $N_c=3$ is close to a 
critical value, $N_c^\ast$, at which the $U(1)_A$ symmetry is restored, 
which  would ``explain'' why lattice simulations
show a sharp drop near $T_c$ even though $N_c$ is only $N_c=3$. 
By the same token, the restoration of $U(1)_A$ could become 
more pronounced closer to the critical point.
    
These arguments suggest that CP-odd false vacua might exist in hot QCD. 
While our assumptions might be wrong, they aren't completely crazy,
at least to us.  If anything, 
we have exhibited an explicit mechanism through 
which CP and P-odd effects could exist in QCD, despite 
the fact that $\theta\approx 0$ to a very high precision in Nature. 
Whether these effects states could have observable consequences 
is another question. Metastable vacua 
such as those discussed here could be formed in the early universe, 
near the QCD phase transition, or in heavy ion collisions. The latter 
possibility will be the topic of the last section.
Of course, the outcome will depend very much on the probability of formation of
these metastable states, of their size and number density and, finally, 
on their lifetime. Once they are allowed to exist, a metastable region
is easily formed assuming that the phase of the 
singlet Goldstone mode, {\em i.e.} $\theta_{eff}$, is 
initially random, as the root mean square of $\theta_{eff}$ is 
quite large, $\langle\theta_{eff}\rangle_{rms} = 2 \pi/\sqrt{3}$. 
The size $l_B$ of a CP-odd bubble is
harder to estimate given the uncertainties of our model, but it 
could be as large as a few fermi. Finally, the lifetime will depends
of the relative size of the false and true vacua region, the difference in 
pressure between these regions (which could be rather small if $U(1)_A$ 
is very nearly restored), and other factors, like the viscosity 
of the medium. All together, there are many uncertainties. The most 
favorable case is
obviously if very large bubbles, of about the size of the hot region,
can be formed.
On the opposite, if many tiny CP-odd regions form 
and quickly collapse, all CP violating effect would average to zero. 
Whether this or the former (or none of the above) is realized is 
a question which, at this stage, can only answered by the experiments.  
	
\section{Parity odd bubbles in heavy ion collisions: the observables}
\noindent

Hot QCD matter can be produced and 
studied in ongoing experiments with relativistic heavy ion collisions. 
It is therefore important 
to investigate the possible signatures associated with the excitation 
of metastable states discussed above. 
From (\ref{pot}), (\ref{ampot}) it is clear that these states will act 
like regions of nonzero $\theta$.  Parity, CP, and time invariance
 will be violated
spontaneously in such a region.  

The first signature of such states stems from the observation that 
when the anomaly term $a$ becomes small, there is maximal violation
of isospin.\cite{largeNphen1,ren1,small1,small2}  At zero temperature, the 
nonet of pseudo-Goldstone bosons --- the $\pi$'s, $K$'s, $\eta$,
and $\eta'$, are, to a good approximation, eigenstates of $SU(3)$
flavor.  It is not often appreciated, but this is really due to
the fact that the anomaly term is large, splitting off the $\eta'$
to be entirely an $SU(3)$ singlet.  When the anomaly term becomes
small, however, while the charged pseudo-Goldstone bosons remain
approximate eigenstates of flavor, the neutral ones do not.
Without the anomaly, the $\pi^0$ becomes pure $\overline{u} u$,
the $\eta$ pure $\overline{d} d$, and the $\eta'$ pure
$\overline{s} s$.  Consequently, these three mesons become light.
This is especially pronounced for the $\eta$, as it sheds all
of its strangeness, to become purely $\overline{d} d$.
Thus the $\eta$ and $\eta'$ would be produced copiously,
and would manifest itself in at least two ways.  First,
light $\eta$'s and $\eta'$'s decay into two photons, and
so produce an excess at low momentum.  Secondly, these
mesons decay into pions, which would be seen in Bose-Einstein
correlations\cite{bose}. Further, through Dalitz decays, 
the enhanced production 
of  $\eta$'s and $\eta'$'s will enhance the yield of low mass dileptons
\cite{small2}. 

Recently, a detailed study of the $\eta$ and $\eta'$ enhancement 
was carried out by 
Ahrensmeier, Baier and Dirks \cite{baier}, who considered the 
dynamical evolution 
of these fields in the decay of the P-odd bubbles. Their result 
is a dramatic enhancement 
in the production of flavor-singlet mesons, leading to about a 
hundred particles 
in a single event. Similar studies, 
%with a different 
%effective potential and somewhat different numerical results, 
were recently presented by 
Buckley, Fugleberg, and Zhitnitsky \cite{bfz}.
 
The maximal violation of isospin is true whenever the anomaly
term becomes small.  There are other signals which only
appear when parity odd bubbles are produced.  Since parity
is spontaneously violated in such a bubble, various decays,
not allowed in the parity symmetric vacuum, are possible.
Most notably, the $\eta$ can decay not just to three pions,
as at zero temperature, but to two pions.  Because of the kinematics,
in a parity odd bubble, $\eta \rightarrow \pi^0 \pi^0$ is allowed,
but $\eta \rightarrow \pi^+ \pi^-$ is not. 
Unfortunately, the search for the parity violating decays will be 
made difficult by the fact that the masses of the $\eta$ and $\eta'$ in the 
bubble are different from their canonical values, and depend on the 
temperature of the system. 

There is another measure of how parity may be 
violated \cite{Kharzeev:1998kz}, which we will 
now discuss following the paper \cite{kp}.  We first
argue by analogy.  Consider propagation in a background magnetic
field.  As charged particles propagate in the magnetic field,
those with positive charge are bent one way, and those with
negative charge, the other.  This could be observed by measuring
the following variable globally, on an event--by--event basis:
\begin{equation}
{\bf J} = \sum_{\pi^+\pi^-}
\frac{(\vec{p}_{\pi^+} \times \vec{p}_{\pi^-})\cdot \hat{z}}
{ |\vec{p}_{\pi^+}| |\vec{p}_{\pi^-}|} \; ;
\label{eq:eh}
\end{equation}
here $\vec{z}$ is the beam axis, and $\vec{p}$ are the three momenta
of the pions.

If the quarks were propagating through a background
chromo-magnetic field, then ${\bf J}$, which is like handedness
in jet physics,\cite{hand} is precisely the right quantity.
However, a parity odd bubble is
not directly analogous to a background chromo--magnetic field: $\pi^+$'s
and $\pi^-$'s propagate in a region with constant but
nonzero $\phi$ in the same fashion.  Consider, however,
the edge of the parity odd bubble: in such a region, 
$U^\dagger \partial_\mu U$ is nonzero, and does rotate $\pi^+$
and $\pi^-$ in opposite directions.  Thus it is the edges
of parity odd bubbles which contribute to the parity
odd asymmetry of (\ref{eq:eh}).  Purely on geometric grounds,
this suggests that a reasonable
estimate for the maximal value of ${\cal P}$ is on the order of
a few percent.

Let us now construct  
global observables which are odd under the discrete symmetries of 
$\cal P$, $\cal C$, and/or $\cal CP$ using only the momenta of 
charged pions in an event.
Our discussion will be general, 
independent of whatever detailed mechanism might produce nonzero 
values for these variables.  For the collisions of nuclei with equal atomic
number, as the initial state is even under $\cal P$, the observation
of a $\cal P$-odd final state must be due to parity violation, such
as by a $\cal P$-odd bubble.  
Based upon our specific model \cite{Kharzeev:1998kz}, we will 
then give a rough estimate of 
the magnitude of the $\cal P$-odd and $\cal CP$-odd effects;
we find that the asymmetries can be relatively
large, at least $\sim 10^{-3}$.

At high energy, nucleus-nucleus collisions produce many pions,
on the order of $\sim 1000$ per unit rapidity at RHIC energies.
Experimentally, it is probably easiest to detect charged pions and their
three-momenta.  (All of our comments apply equally well to charged
kaons.)  Thus we are led to consider constructing observables
only from the three-momenta of $\pi^+$'s, $\vec{p}_+$, and 
$\pi^-$'s, $\vec{p}_-$.  As vectors, under parity the three-momenta 
transform as 
\begin{equation}
{\cal P}: \;\;\; \vec{p}_+ \rightarrow - \vec{p}_+ \;\;\; , \;\;\; 
\vec{p}_- \rightarrow - \vec{p}_- \; .
\end{equation}
Charge conjugation switches 
$\pi^+$ and $\pi^-$, 
\begin{equation}
{\cal C}: \;\;\; \vec{p}_+ \leftrightarrow \vec{p}_-  .
\end{equation}
Lastly, time reversal, ${\cal T}$, acts like parity on the
pion momenta, switching their sign:
$\vec{p}_+ \rightarrow - \vec{p}_+$ and $\vec{p}_- \rightarrow - \vec{p}_-$
under ${\cal T}$.

It is a theorem that any $\cal P$-odd invariant 
formed from three-vectors can be represented as a sum of terms,
each of which involves one antisymmetric epsilon tensor \cite{weyl}.
The variable which we considered previously (\ref{eq:eh}) 
is of this type \cite{Kharzeev:1998kz}.
In order to form ${\bf J}$ we have introduced an arbitrary, fixed
vector of unit norm, $\hat{z}$.  If $\hat{z} \rightarrow - \hat{z}$
under parity, then ${\bf J}$ is odd under ${\cal P}$.
In ${\bf J}$ we use the unit vectors 
$\hat{p}_\pm = \vec{p}_\pm/|\vec{p}_\pm|$ so that
it is a pure, dimensionless number.
The variable ${\bf J}$ is separately odd under $\cal P$ and $\cal C$,
and so is even under $\cal CP$.  It is even under $\cal T$
and $\cal CPT$.

The variable $\bf J$ is closely analogous to ``handedness'', originally
introduced to study spin dependent effects in jet
fragmentation \cite{hand}.
There the axis $\hat{z}$ is usually defined by
the thrust of the jet, with $\hat{p}_+$ and $\hat{p}_-$ representing
the directions of pions formed in the fragmentation of the jet.
Correlations between the handedness of different jets produced
in a given event are sensitive to $\cal CP$-violating effects \cite{hand}.

It is not difficult to construct
other invariants with different transformation properties.  We
introduce $\vec{k}_\pm$ as
\begin{equation}
\vec{k}_\pm = \sum_{\pi^+} \vec{p}_+ \pm \sum_{\pi^-}\vec{p}_- \;\;\; , \;\;\;
\hat{k}_\pm = \vec{k}_\pm/k_\pm \; , 
\end{equation}
and then form
\begin{equation}
{\bf K}_\pm = 
\sum_{\pi^+, \pi^-}
(\hat{p}_{+} \times \hat{p}_{-})\cdot \hat{k}_\pm \; .
\label{eqk}
\end{equation}
The variables ${\bf K}_\pm$ are $\cal P$-odd;
${\bf K}_+$ is $\cal C$-odd, and so $\cal CP$-even,
while ${\bf K}_-$ is $\cal C$-even, and so $\cal CP$-odd.
Both ${\bf K}_\pm$ are $\cal T$-odd, so that ${\bf K}_+$ is
$\cal CPT$-odd, and ${\cal K}_-$ is $\cal CPT$-even.
The vector $\vec{k}_+$ measures the net flow of the charged pion momentum,
while $\vec{k}_-$ measures the net flow of charge from pions.

We can also form
\begin{equation}
{\bf L} = 
\sum_{\pi^+, \pi^-}
(\hat{p}_{+} \times \hat{p}_{-})\cdot \hat{z} 
\; \; \left( \frac{p_+^2 - p_-^2}{p_+^2 + p_-^2} \right) \; .
\end{equation}
The variable $\bf L$ is $\cal P$-odd,
$\cal C$-even, and $\cal T$-even, so it is odd under
$\cal CP$ and $\cal CPT$.
It does not require a net flow
of momentum or charge to be nonzero, although as for $\bf J$, we do need
to introduce an arbitrary unit vector $\hat{z}$.

Similar to $\bf L$, we can form \cite{altesun}
\begin{equation}
{\bf M} = 
\sum_{\pi^+, \pi^-}
\; \; \left( \frac{p_+^2 - p_-^2}{p_+^2 + p_-^2} \right) \; .
\end{equation}
This variable is $\cal P$-even, $\cal C$-odd, 
$\cal T$-even, and so odd under $\cal CP$ and $\cal CPT$.

Besides using the vectors $\vec{k}_\pm$, another way of avoiding
the introduction of an arbitrary unit vector $\hat{z}$ is to 
use ordered pairs of pion momenta \cite{lee,gyulassy};
this procedure is also used in the studies of spin-dependent
effects in jet fragmentation \cite{hand}.  For any given
pair of like sign pions, let $\vec{p}^{\; \prime}_+$ denote the $\pi^+$ with
largest momentum, so $|\vec{p}^{\; \prime}_+| > |\vec{p}_+|$.  This ordering
is done in the frame in which the observable is measured.
Then we can form a triple product as
\begin{equation}
{\bf T}_\pm = 
\sum_{\pi^+, \pi^-}
(\hat{p}_{+} \times \hat{p}_{-})\cdot (\hat{p}^{\, \prime}_{+} 
\pm \hat{p}^{\, \prime}_{-}) \;\;\; ,
\;\;\; |\vec{p}^{\; \prime}_\pm| > |\vec{p}_\pm|
\end{equation}
The variables ${\bf T}_\pm$ are odd under $\cal P$ and $\cal T$;
${\bf T}_+$ is $\cal C$-odd, $\cal CP$-even, and $\cal CPT$-odd, 
while ${\bf T}_-$ is $\cal C$-even, $\cal CP$-odd, and $\cal CPT$-even.
Besides the variables ${\bf T}_\pm$, one can clearly construct
other $\cal P$- and $\cal C$-odd observables from triplets,
or higher numbers, of pions.

On general grounds, any scalar observable should be invariant under
the combined operation of $\cal CPT$ \cite{weinberg}.  Consequently,
the $\cal CP$-odd
variables $\bf J$, ${\bf K}_-$, and ${\bf T}_-$ are allowed under
$\cal CPT$,
while ${\bf K}_+$, ${\bf L}$, ${\bf M}$, and ${\bf T}_+$ must vanish.

We have previously derived the metastable $\cal P$-odd bubbles 
within the context of a nonlinear sigma model, with 
a $U(3)$ matrix $U$, $U^\dagger U = \bf 1$.  
The metastable vacua are stationary points with respect to the
nonlinear sigma model action, including the terms with two derivatives,
a mass term, and an anomaly term (\ref{pot}).\
In terms of the underlying gluonic fields, the $\cal P$-odd bubbles
arise from fluctuations in the topological charge density, $G_{\mu \nu}
\widetilde{G}^{\mu \nu}$.  It is easy to understand how
a region in which $G_{\mu \nu} \widetilde{G}^{\mu \nu} \neq 0$
can produce a $\cal P$-odd effect.  Consider the propagation of
a quark anti-quark pair through a regioin in which 
$G_{\mu \nu} \widetilde{G}^{\mu \nu} \neq 0$; in terms of
the color electric, $\vec{E}$, and color magnetic, $\vec{B}$,
fields, $G_{\mu \nu} \widetilde{G}^{\mu \nu} \sim \vec{E} \cdot \vec{B}$.
If $\vec{E}$ and $\vec{B}$ both lie along the $\hat{z}$ direction,
then a quark is bent one way, the anti-quark the other,
so that $(\vec{p}_q \times \vec{p}_{\overline{q}}) \cdot \hat{z} \neq 0$,
where $\vec{p}_q$ and $\vec{p}_{\overline{q}}$ are the three-momenta
of the quark and anti-quark, respectively.  While physically
intuitive, this picture does not allow us to directly relate
this bending in the momenta of the quark and anti-quark to an asymmetry
for charged pions.  

To do so, we again resort to using an effective lagrangian.
It is known that the effects of the axial anomaly show up
in the effective lagrangians of Goldstone bosons in two,
and only two \cite{dhoker}, ways.
One is through the anomaly
term, 
$\sim a$, which we have already included.  Besides that,
however, the axial anomaly also manifests itself in the interactions of
Goldstone bosons through the Wess-Zumino-Witten term \cite{wz,wittenwz}.
This term is nonzero only when the fields are time dependent,
which is why we could ignore it in discussing the static properties
of $\cal P$-odd bubbles.  It cannot be ignored, however, in 
computing the dynamical properties, and in particular the decay,
of $\cal P$-odd bubbles.
The Wess-Zumino-Witten term is manifestly chirally symmetric
when written as an integral over five dimensions,
\begin{equation}
{\cal S}_{wzw} = 
- i {1\over 80 \pi^2} \int d^5 \!x \;
\varepsilon^{\alpha \beta \gamma \delta \sigma}\,
tr\left(R_\alpha R_\beta R_\gamma R_\delta R_\sigma 
\right ) \; , 
\end{equation}
$R_\alpha = U^\dagger \partial_\alpha U $,
but reduces to a boundary 
term in four space-time dimensions.  For $U = exp( i u)$,
when $\partial_\alpha u \ll 1$,
\begin{equation}
{\cal S}_{wzw} \approx
{2\over 5 \pi^2} \int d^4 \!x \;
\varepsilon^{\alpha \beta \gamma \delta }\,
tr\left(u \; \partial_\alpha u \; 
\partial_\beta u \; \partial_\gamma u \;
\partial_\delta u
\right ) \; .
\label{wzw1}
\end{equation}
As discussed by Witten \cite{wittenwz}, the coefficient of
the Wess-Zumino-Witten term is fixed by topological considerations,
and is
proportional to the number of colors, which equals three.

In a collision, we envision that the trivial vacuum heats up,
a $\cal P$-odd bubble forms, and then decays as the vacuum cools.
Since this represents bubble formation and decay, there is no
net change in any topological number.  Therefore, it is possible for
a given event to contain an excess of bubbles over anti-bubbles
(or vice versa), and thus to manifest true parity violation on an
event by event basis.

To estimate the magnitude of the Wess-Zumino-Witten term for
a $\cal P$-odd bubble, and to understand its effect on pion
production, in (\ref{wzw1}) we 
can take three $u$'s to be condensate fields, $u \sim \phi_{u,d,s}$,
and two to be charged pion fields, $u \sim \pi_{\pm}/f_\pi$.  
Suppose that the 
$\cal P$-odd bubble is of size $R$, with unit normal
$\hat{r}$ to the surface, and lasts for some period of time.
Because of the antisymmetric tensor in
(\ref{wzw1}), all three components of the condensate field
must enter.  Schematically, we obtain
\begin{equation}
{\cal S}_{wzw} \approx
{2\over 5 \pi^2} \;
\int dt \int d^3 r  \; \phi_u \; \partial_r \phi_d \; \partial_0 \phi_s \;
(\vec{p}_{\pi^+} \times \vec{p}_{\pi^-}) \cdot \hat{r}
\end{equation}
The time integral is 
$\int dt \, \partial_0 \phi_s \sim \delta \phi_s = \phi_s$, since 
$\phi_s = 0$ in the
normal vacuum.  Similarly, the spatial integral is
$\int d^3r \, \partial_r \phi_d \, (\vec{p}_{\pi^+} \times \vec{p}_{\pi^-}) 
\cdot \hat{r}
\sim \int d\Omega \, \int R^2 dr \, \partial_r \phi_d \,
(\vec{p}_{\pi^+} \times \vec{p}_{\pi^-}) 
\cdot \hat{r} 
\sim \int d\Omega \, R^2 \phi_d \,
(\vec{p}_{\pi^+} \times \vec{p}_{\pi^-}) 
\cdot \hat{r}$, where $\int d\Omega$ represents an
integral over the direction of the normal, $\hat{r}$.  
Further, as the average momentum within the condensate is
of order $p_\pi \sim 1/R$, the size of the bubble drops out as well.
We thus obtain a final result which is independent of the size
of the bubble, its lifetime, and its width:
\begin{equation}
{\cal S}_{wzw} \approx
\frac{2 \phi_u \phi_d \phi_s}{5 \pi^2} \int d \Omega \;
(\hat{p}_{\pi^+} \times \hat{p}_{\pi^-}) \cdot \hat{r}
\label{wzw2}
\end{equation}
Since the Wess-Zumino-Witten term vanishes for a static field,
an asymmetry is only obtained from the decay of a $\cal P$-odd
bubble.  In addition, scattering of pions off the $\cal P$-odd
bubble will also produce an asymmetry, although we do not
include this at present.  Lastly, note that there is only
an observable asymmetry when $\phi_s \neq 0$; this is because
in the absence of external gauge fields, there is only a Wess-Zumino-Witten
term for three, and not for two, flavors.  Within this model,
${\cal S}_{wzw}$ is of similar form for two charged kaons.

Using our previous estimates for the $\phi$'s, $\phi_u \sim \phi_d \sim 1$
and $\phi_s \sim 10^{-2}$, we obtain an effect of order
$\sim 10^{-3}$.
At the point where the $\cal P$-odd bubble first appears,
$(a/c)_{cr}$, one can estimate that the energy density
within the bubble, relative to the ordinary
vacuum, is $\sim 25 \, n^2 \, MeV/fm^3$, where
$n$ is the winding number of the bubble, $n=1,2,3...$
For a bubble $\sim 5 fm$ in radius, there are about
$\sim 100 \, n^2$ pions produced in the decay of the bubble.
If a fraction of the produced pions are observed within a
given kinematical window, and we assume that 
all observed pions come from a portion of the total bubble,
then we recover the variable $\bf J$, introduced before in
(\ref{eq:eh}), and find an estimate of ${\bf J} \sim 10^{-3}$.  
Moreover, we find a natural interpretation of the direction
$\hat{z}$, which was needed to define $\bf J$, as the normal
to the bubble's surface.  One might wonder if the effect is
diluted by the necessity to average over uncorrelated pairs.
This does not happen, however, because the pion field within
the bubble is a classical field, so that all charged pions are
affected similarly.  

Naively, one might expect that $\bf J$ would average to zero over
a single bubble.  As the bubble is topological, though, the
direction in which charged pions are swept is correlated with
the sign of the condensate, so that a single $\cal P$-odd bubble
can produce an effect in ${\bf J} \sim 10^{-3}$.  
Thus it is possible to distinguish between events in which bubbles are
produced, and those in which bubbles are not, by measuring
$\bf J$.

At first it may seem surprising that our $\cal P$-odd,
$\cal C$-even, and $\cal CP$-odd bubble
produces a signal in $\bf J$, which by previous analysis is
$\cal P$-odd, $\cal C$-odd, and $\cal CP$-even.  
The Wess-Zumino-Witten term is even under parity, which
is ${\cal P}_0 (-1)^{N_B}$, where ${\cal P}_0$ is the operation
of space reflection, and $N_B$ counts the number of 
Goldstone bosons \cite{wittenwz}.  By
scattering off a $\cal P$-odd bubble, we bring in
an odd number of condensate fields, ${\bf J} \sim \phi_u \phi_d \phi_s$,
(\ref{wzw2}), 
and so change the (apparent) quantum numbers to be $\cal P$-odd and
$\cal CP$-even.
This is only apparent, as scattering off an anti-bubble will give
the opposite sign of $\bf J$.

We expect that bubbles will generate signals for the other
variables presented of similar magnitude.  For example, a
single bubble will induce a 
net flow of pion charge, and so contribute to
${\bf K}_- \sim 10^{-3}$, (\ref{eqk}).  Through coherent
scattering in a bubble, we would also expect the variables
${\bf K}_+$, $\bf L$, and $T_\pm$ to develop signals
$\sim 10^{-3}$. Further, hot gauge
theories can also exhibit metastable states which are 
$\cal P$-even and $\cal C$-odd
\cite{altes}; these generate signals for $\bf M$ \cite{altesun}.

The idea of exciting metastable vacua in hadronic reactions
is an old one \cite{leewick,Morley:1985wr}, as is the idea that
a collective pion field can produce large fluctuations
in heavy ion collisions on an event by event basis
\cite{dcc}. We wish to emphasize that the $\theta$ dependence 
in QCD, and associated with it complicated vacuum structure, 
may lead to the possibility of exciting the metastable 
vacuum states with broken 
discrete symmetries of P, CP, and T in experiments with 
relativistic heavy ions. 
The search for these phenomena at RHIC is underway, 
and we are anxiously awaiting 
the experimental verdict.

\nonumsection{Acknowledgements}
\noindent 
We are grateful to Chris Korthals-Altes for his invitation to SEWM2000 and 
kind hospitality in Marseille. We would like to thank 
Misha Stephanov for discussions. One of us (M.T.) thanks the organizers of ISMD2000 
for a very pleasant meeting.

\nonumsection{References}

\end{document}